\documentclass{elsart}
\usepackage{graphicx}
\begin{document}
\begin{frontmatter}

\title{Lattice-Boltzmann Method for Geophysical Plastic Flows} 

\author{Alessandro Leonardi}
\author{Falk K. Wittel} 
\author{Miller Mendoza} 
\author{Hans J. Herrmann} 

\address{ETH Zurich, Institute for Building Materials, Schafmattstrasse 6, 8093, Zurich CH} 

 \begin{abstract}
We explore possible applications of the Lattice-Boltzmann Method for the simulation of geophysical flows. This fluid solver, while successful in other fields, is still rarely used for geotechnical applications. We show how the standard method can be modified to represent free-surface realization of mudflows, debris flows, and in general any plastic flow, through the implementation of a Bingham constitutive model. The chapter is completed by an example of a full-scale simulation of a plastic fluid flowing down an inclined channel and depositing on a flat surface. An application is given, where the fluid interacts with a vertical obstacle in the channel.
\end{abstract}
\end{frontmatter}

\newpage

\section{Introduction}
\label{introduction}
Geophysical flows are dangerous natural hazards occurring mostly in mountainous terrain. The most apparent phenomena of this category are debris flows, which originate when heavy rainfall mobilizes a large amount of debris \cite{Iverson1997}. The resulting mixture comprises water, cohesive sediments, organic matter, silt, sand and in many cases also stones of different sizes The resulting rheological behavior is known to have a wide variability\cite{Geographie}, which makes numerical studies an essential tool to support experimental investigations \cite{Hutter1996}.
Full-scale simulations of geophysical flows are very scarce, since they require a framework that efficiently manages complicated boundary conditions, as well as a powerful and flexible fluid solver. Moreover, the non-Newtonian nature of the material, and in some cases its multiple phases, pose more challenges. Traditional solvers are known to have troubles in tackling this problem, and nowadays alternative solution are sought by the community.

The Lattice Boltzmann Method (LBM) \cite{Succi2001} is becoming increasingly popular and is today considered a valid alternative for categories of flows where traditional solvers exhibit disadvantages, like multiphase fluids, flows through porous media \cite{Mendoza2010a}, irregular geometries \cite{Mendoza2013}, and free-surface realizations \cite{Korner2005}.

After reviewing the most commonly used rheological model for flowing geomaterials in Sec. \ref{rheology}, we offer an essential overview of the method in Sec. \ref{LBM}, together with a simple but effective formulation for the simulation of geophysical flows. In Sec. \ref{simulation} and \ref{obstacle} examples are given.

\section{Rheology of geophysical flows and simulation}
\label{rheology}

The rheology of geophysical flow materials is a debated issue in the field, due to the extreme variability in natural material parameters and the presence of multiple phases, complicating the classification. Most models therefore adopt simplified solutions based on single-phase descriptions. This can either be a frictional material \cite{Savage1989,Mcdougall2004} or a viscoplastic fluid \cite{Dent1983,Whipple1997}. The former is used for rock and snow avalanches, while the latter is preferred for mudflows and viscous debris flows \cite{Ancey2007}. For certain categories of geophysical flows, however, a single-phase approach is insufficient to capture the physics of the phenomena. Debris flows are a typical example of this, because granular and viscous behavior interact, giving rise to unexpected structures and a localization of rheological properties \cite{Iverson}. A continuum-continuum coupling for granular and fluid phase is possible, but is incapable of capturing the localization of flow properties, 
which 
is widely recognized to be a key feature of debris flows.

A discrete-continuum approach would of course be able to provide a detailed description, but development of this sort of coupling has been slowed by its demanding computational cost. This is currently challenged, however, by the maturity reached by alternative solvers like Smoothed Particle Hydrodynamics, the Material Point Method, or LBM, which are more flexible in managing complex boundary conditions than traditional tools like Finite Differences and Finite Volumes. In such methods, the granular phase is treated by a separate solver and, for this reason, the fluid model can focus on the nature of the material. This is the reason behind our choice to adopt LBM with a purely viscoplastic rheological law, an approach that can offer:
\begin{itemize}
 \item An efficient framework for the simulation of geophysical flows of plastic nature, where the complexity of the boundary does not influence the performance.
 \item A convenient environment for the coupling with a discrete method. This option opens future chances for full realizations of multiphase flows \cite{leonardiSubmitted}.
\end{itemize}

Regarding the specific rheological law, we adopt the Bingham model, which is widely used to describe plastic fluids due to its conceptual simplicity. It reads:
\begin{equation}
\label{eqBingham}
   \left\{
  \begin{array}{l l}
    \dot{\gamma}=0 & \quad \textrm{if fluid does not yield}\ (\sigma<\sigma_y),\\
    \sigma = \sigma_y +\mu_{pl} \dot{\gamma} & \quad \textrm{if fluid flows}\ (\sigma>\sigma_y),\\
  \end{array} \right.\
\end{equation}
where $\sigma_y$ and $\mu_{pl}$ denote yield stress and plastic viscosity.  An analogous way to write the law is through an analogy with Newtonian flow. One defines a parameter, the apparent viscosity $\mu_{app}$, which proportionally relates stress and rate of strain and is treated as a variable. In the case of a Bingham fluid,  $\mu_{app}$ takes the form
\begin{equation}
\label{eqBinghamMu}
\sigma =\mu_{app} \dot{\gamma} \Rightarrow \mu_{app} = \mu_{pl}+\frac{\sigma_y}{\dot{\gamma}},
\end{equation}
where the apparent viscosity $\mu_{app}$ (from now on, for simplicity, called viscosity $\mu$), diverges when $\dot{\gamma} \rightarrow 0$, which will require special care in the solver.
We are now ready to introduce LBM in the next section, and to incorporate this constitutive law in Sec. \ref{NN}.

\section{Lattice-Boltzmann Formulation}
\label{LBM}

LBM has lately emerged as an attractive alternative to traditional fluid solvers, mainly due to its high-level performance and the predisposition to parallelization. LBM is also suitable to the solution of problems involving complex boundary conditions  \cite{Aidun2010}. It is beyond the scope of this chapter to give a complete description of the method. The reader can refer to Refs. \cite{Chen1998,Succi2001} for a comprehensive review. We will focus on the aspects of the formulation that need to be modified in order to successfully reproduce debris flows.

In LBM, the fluid is described using a distribution function $f_i$ and a set of discrete velocities $\mathbf{c}_i$. Density $\rho$ and velocity $\mathbf{u}$ of the fluid are computed as the first two moments of the distribution function
\begin{eqnarray}
\label{densityVelocity}
 \rho=\sum\limits_i f_i, &
 \mathbf{u}=\sum\limits_i f_i\mathbf{c}_i/\rho.
\end{eqnarray}
The evolution of $f_i$ is governed by the Lattice-Boltzmann equation
\begin{eqnarray}
\label{newfunctions}
f_i (\mathbf{x}+ \delta t \mathbf{c}_i,t+\delta t)= f_i (\mathbf{x},t)+\Omega_i (\mathbf{x},t) ,
\end{eqnarray}
where $\Omega_i$ is the operator that represents the effects of inter-particle collisions in the fluid.
A common way to approximate the otherwise complex expression of $\Omega_i$ is the Bhatnagar-Gross-Krook operator \cite{Bhatnagar1954}, which relaxes the distribution function to a thermodynamic equilibrium $f^{eq}_i$. It can be written as
\begin{eqnarray} \label{collision}
\Omega_i=\delta t \left( \frac{f_i^{eq}-f_i}{\tau} \right),
\end{eqnarray}
and features a constant, the relaxation time $\tau$, which is related to the kinematic viscosity of the fluid $\mu$ as
\begin{eqnarray}
\label{mu}
\tau=\frac{\delta t}{2}+\frac{\mu}{c_s^2}.
\end{eqnarray}
With this formulation, and with the setting of a coherent lattice \cite{Shan1998}, LBM can produce realizations of fluid dynamics in analogy to the Navier-Stokes equations. The method is accurate in the limit of small Mach number, practically $\mathbf{u}_{max}<0.01 c_s$ with $c_s$ denoting the lattice speed of sound.
We will now describe two additions to the model necessary for the simulation of geophysical flows: a non-Newtonian rheology and a free-surface treatment.

\subsection{Non-Newtonian rheology}
\label{NN}
\begin{figure}[t]
\centering
\includegraphics[width=0.6\textwidth]{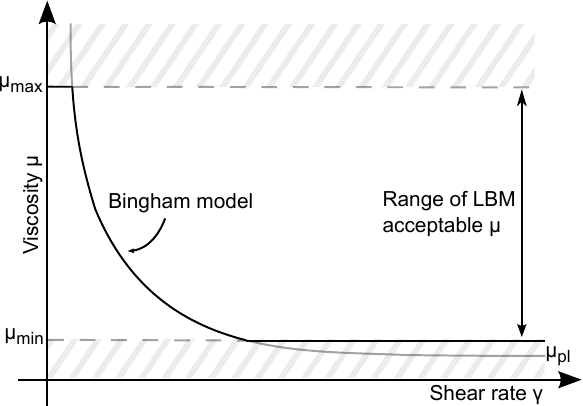}
\caption{Representation of the rheology model employed for plastic fluids. The approximation of the Bingham model is limited by the maximum and minimum values for the relaxation time $\tau$ imposed by the method. Therefore, also the viscosity $\mu_f$ is limited.}
\label{trilinear}
\end{figure}
The LBM described in the previous section yields, after the Chapman-Enskog expansion \cite{chapman1970mathematical}, the Navier-Stokes equation for Newtonian fluids. A simple way to upgrade the method to more general formulations is offered by a local treatment of the relaxation time $\tau$ \cite{Leonardi2011,Svec2012}. Any rheological law that can be approximated as
\begin{equation}
 \sigma = \mu\dot{ \gamma},
\end{equation}
with $\mu=\mu (\dot{\gamma})$, is suitable for this approach. The relaxation time can in fact be directly related to the viscosity through Equation \ref{mu}, obtaining ad hoc formulations for different rheological laws. The Bingham fluid, for example, can be written as
\begin{equation}
\label{bingham}
 \sigma=\sigma_y+\mu_{pl}\dot{ \gamma} \Rightarrow \tau=
 \frac{\delta t}{2}+\frac{1}{c_s^2}\left(\mu_{pl}+\frac{\sigma_y}{\dot{ \gamma}}\right).
\end{equation}
This type of formulation requires the computation of the shear rate tensor, which can be done easily in LBM directly form the distribution functions
\begin{eqnarray}
 \dot{\gamma}_{ab} = \frac{1}{2\tau c_s^2}
 \sum_i \mathbf{c}_{i,a} \mathbf{c}_{i,b}
 \left( f_i  -f_i^{eq}\right),
\end{eqnarray}
and the magnitude can be extracted as
\begin{equation}
\label{gamma}
  \dot{\gamma}=\sqrt{2 \sum_a \sum_b  \dot{\gamma}_{ab}\dot{\gamma}_{ab}}.
\end{equation}
The limitation of this approach lies in the range of values given to the relaxation time $\tau$ by Equation \ref{bingham}.  Accuracy in LBM is guaranteed as long as $\tau_{min}<\tau<\tau_{max}$. Reasonable values for these limits are $\tau_{min}=0.501$ and $\tau_{max}=1.0$. Therefore, also the viscosity $\mu$, which is linearly linked to the relaxation time, is subjected to the same restrictions: $\mu_{min}<\mu<\mu_{max}$. The following considerations are thus necessary:
\begin{itemize}
 \item The fluid that reaches the maximum allowed value of $\mu$ is considered to be in a plastic state. However, with the proposed scheme, the fluid never stops its motion, but rather flows at a much slower rate. The ratio between $\mu_{max}$ and $\mu_{min}$ determines the effectiveness of this approach. With the proposed limit values for $\tau$, $\mu_{max}=500\mu_{min}$.
 \item  The best approximation of a Bingham fluid is obtained when $\mu_{min} \leq \mu_{pl}$, because the lower limitation on $\mu$ has no effect. However, an eventual transition to turbulent regime can happen when simulating diluted flows, and therefore the value of $\mu_{min}$ must be raised to avoid instabilities. In case  $\mu_{min} \geq \mu_{pl}$, the approximation of the Bingham constitutive model becomes less accurate.
\end{itemize}

\subsection{Implementation of the free-surface technique}

In order to simulate geophysical flows on realistic geometries, we need to include the boundary conditions given by the channel bed and the interface of the flow with air. While the former can be implemented as a standard no-slip boundary condition, as in Ref. \cite{Mohamad2011}, the latter is a less common practice in LBM. The free-surface is represented through a classification of the lattice nodes in three categories: liquid, interface and gas nodes. The governing parameter is the liquid fraction $\lambda$:
\begin{eqnarray}
\left\{
  \begin{array}{l l}
    \lambda=\rho & \quad \textrm{if the node is liquid,}\\
    0<\lambda<\rho & \quad \textrm{if the node is interface,}\\
    \lambda=0 & \quad \textrm{if the node is gas.}\\
  \end{array} \right.\
\end{eqnarray}
The liquid fraction of a node evolves according to the streaming of the distribution function given by Equation \ref{newfunctions} as
\begin{eqnarray}
 \lambda(t+\delta t) = \lambda (t) + \frac{\delta t}{\rho}\sum \alpha \left( f_{in} - f_{out} \right)
\end{eqnarray}
where $f_{in}$ and $ f_{out}$ represent the distribution function streaming respectively in and out of the node, and $\alpha$ is a parameter that depends on whether the distributions are exchanged with a fluid node or another interface node. This method conserves mass exactly, and ensures a smooth evolution of the surface. Further details are found in Ref. \cite{Korner2005}.

\section{Simulation of mudflow}
\label{simulation}
\begin{figure}[htb]
\centering
\includegraphics[width=\textwidth]{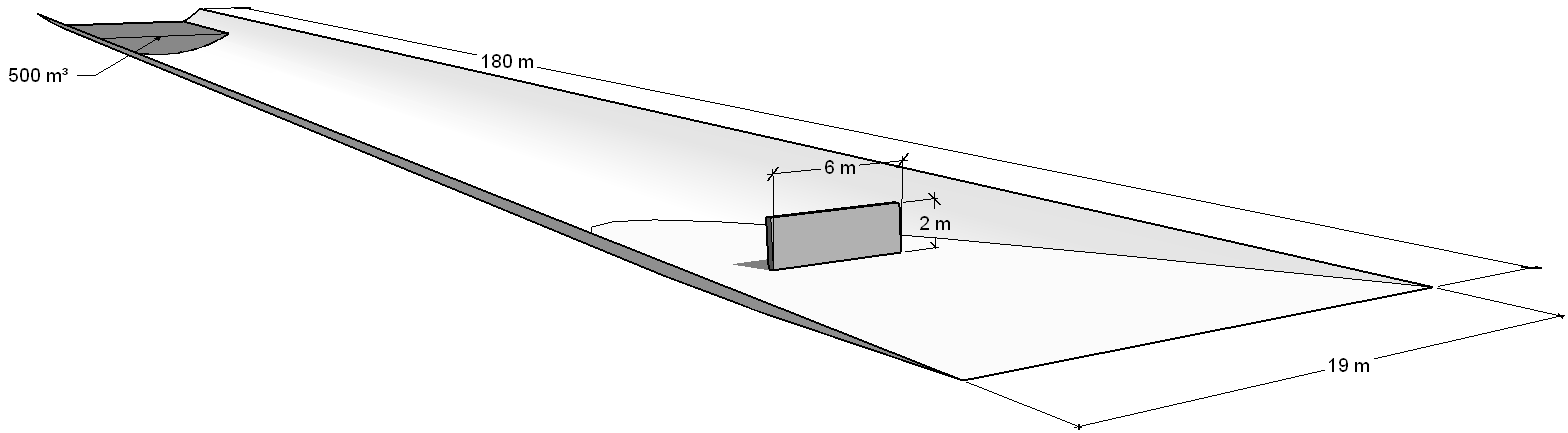}
\caption{Geometry of the simulation. The fluid mass lies at the top of a long cylindrical chute. The deposition area at the bottom is flat and features a vertical obstacle.}
\label{geometry}
\end{figure}
\begin{figure}[htb]
\centering
\includegraphics[]{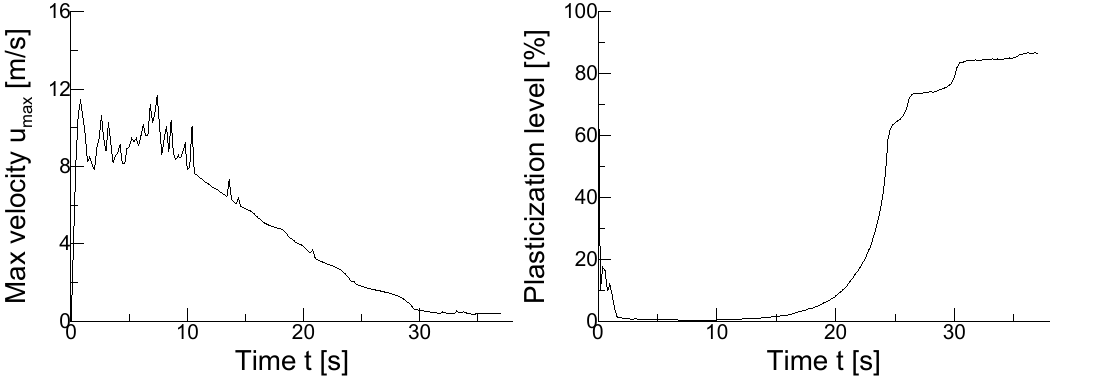}
\caption{Evolution of the maximum velocity of the fluid and of the flow plasticization level, computed as ratio between cells that have reached the maximum viscosity and the total number of cells.}
\label{characteristics}
\end{figure}
\begin{figure}[htb]
\centering
\includegraphics[]{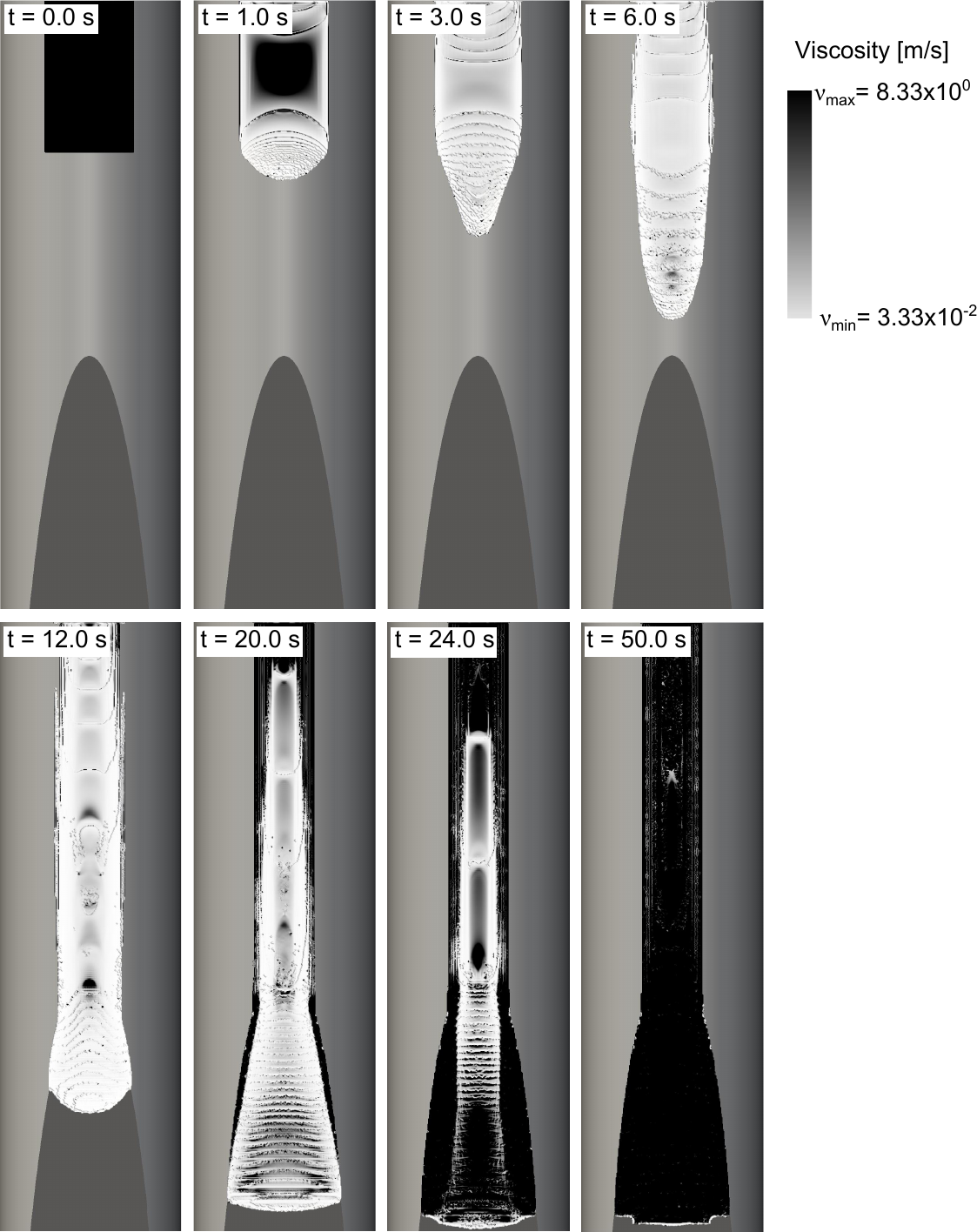}
\caption{Evolution of the geometry of the flow. Intensities show the viscosity at the surface, therefore indicating the rate of shearing of the fluid: low (dark) or high (light).}
\label{flowVisualization}
\end{figure}

The full-scale simulation of a plastic geophysical flow is shown in this section. Mimicking the real geometry of a small valley, the simulation features a cylindrical channel inclined at $5^{\circ}$ with respect to the horizontal, and a flat deposition area at its bottom (Fig. \ref{geometry}). The total volume of the flowing material is $500 \ \textrm{m}^3$ and is fixed, i.e. neither entrainment nor deposition are modeled. While very big events can be of the order of $10^6 \ \textrm{m}^3$, the size of the most frequent type of geophysical flows lies in the range of $10^3 \ \textrm{m}^3$, which is big enough to endanger humans and infrastructures. Therefore our simulation proposes a realistic scenario, even though not a particularly dangerous one.  The fluid has density $\rho = 2000 \ \textrm{kg}/\textrm{m}^3$ and follows a Bingham-like rheological law, like the one proposed in Sec. \ref{NN}. Yield stress and kinematic plastic viscosity are respectively $\sigma_y = 150 \ \textrm{Pa}$ and $\mu_{pl}= 10 \ \
textrm{m}
^2/\textrm{s}$, relating the simulated system to a very dense mudflow or to a debris
flow whose granular phase has been homogenized into the fluid, therefore increasing the bulk viscosity \cite{Phillips1991,coussot1997mudflow}.
Fig. \ref{flowVisualization} shows how the fluid free surface evolves in time. The fluid is quickly sheared by the effect of gravity and moves until an equilibrium is reached in the deposition area, where the viscosity increases. This technique can be used to estimate the deposition area of the material after an event, and to  support the design of hazard maps on real terrain. Fig. \ref{characteristics} shows the evolution of the maximum velocity in the fluid and of the plasticization level of the material, which are the useful parameters to determine the status of the flow.

\section{Obstacle Interaction}
\label{obstacle}
\begin{figure}[htb]
\centering
\includegraphics[]{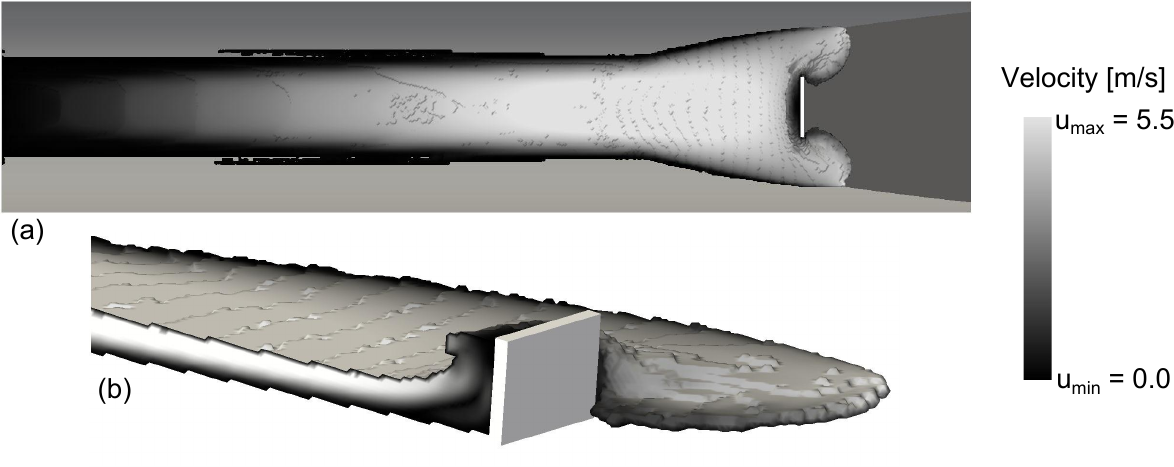}
\caption{Image of the flow splashing on the retaining wall, at $t=14.0 \ \textrm{s}$. The color contour shows the velocity at the free surface (a) and in the longitudinal section (b).}
\label{obstacleImage}
\end{figure}
\begin{figure}[htb]
\centering
\includegraphics[]{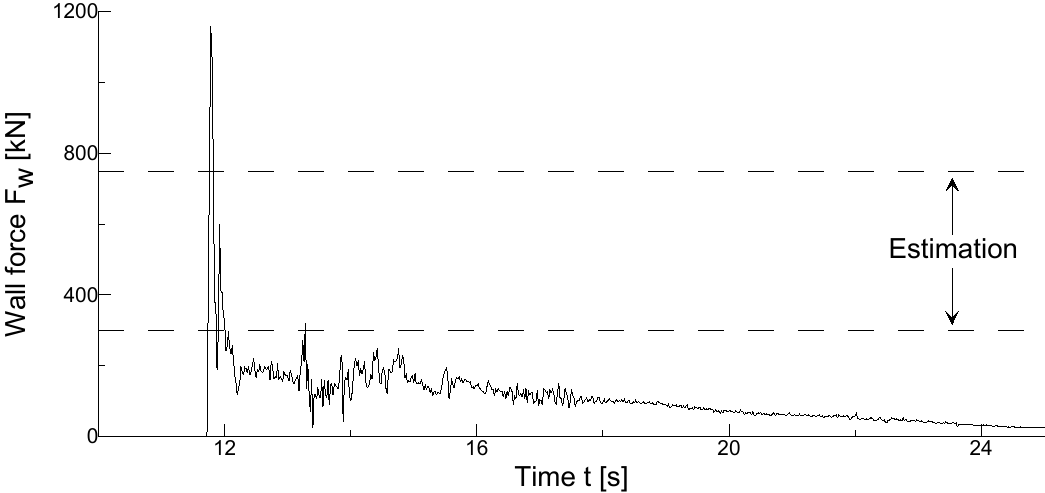}
\caption{Force exerted on the obstacle by the flow. The estimation is obtained with the hydrodynamic formula in Eq. \ref{F_Estimation}}.
\label{forceImage}
\end{figure}
To show the possibilities to use LBM to design protection structures, we repeat the simulation of the previous section, this time featuring an obstacle. LBM can in fact be used to calculate the hydrodynamic interactions on solid objects, computing all momentum transfers between the distribution function and the solid boundaries. The procedure, which is found in Ref. \cite{Li2004}, does not change significantly the overall efficiency of the method.We add a retaining wall, fixed at the bottom of the channel and of size $H \times L \times S = 3.0 \ \textrm{m} \ \times 2.0 \ \textrm{m} \ \times 0.25 \ \textrm{m}$, as in Fig. \ref{geometry}. The shape of the free surface after the impact is shown in Fig. \ref{obstacleImage}, with insight into the longitudinal cross section of the flow.

The force on the wall can be estimated with a hydrodynamic formula \cite{Hubl2009} as
\begin{eqnarray}
 F_{wall}=k A \rho v_{front}^2,
 \label{F_Estimation}
\end{eqnarray}
where $A$ is the area of the obstacle impacted by the flow. The value of the coefficient $k$ is given by the comparison with experiments and varies, according to different authors, from $2$ to $5$. In the simulation, when the flow hits the wall, the depth is $0.5 \textrm{m}$ and the front speed is $v_{front} \simeq 5 \textrm{m}/\textrm{s}$, which leads to an estimated force of $F_{wall} = 300 \div 750 \ \textrm{kN}$. Fig. \ref{forceImage} shows the hydrodynamic force as calculated by the solver, highlighting the importance of the dynamic load due to the initial impact. The maximum values match the prediction of the hydrodynamic formula.

\section{Outlook}
In this chapter we showed how a model based on LBM can be used to simulate geophysical flows and provide a new tool for the rational design of mitigation and protection structures. The model inherits the advantages of the local solution mechanism of LBM, and extends the standard solver with the addition of a Bingham fluid formulation and of the free-surface technique. The resulting framework can be used to simulate homogeneous plastic flows, and provides an optimal environment for the coupling with discrete method, thus opening future chances for the full simulation of multiphase geophysical flows.

\section{Acknowledgements}
The research leading to these results has received funding from the European Union (FP7/2007-2013) under grant agreement n. 289911. We acknowledge financial support from the European Research Council (ERC) Advanced Grant 319968-FlowCCS. The authors are grateful for the support of the European research network MUMOLADE (Multiscale Modelling of Landslides and Debris Flows).

\end{document}